\documentclass[aps,nofootinbib,prd,superscriptaddress,tightenlines,notitlepage,twocolumn,showpacs]{revtex4-1}

\usepackage[colorlinks=true, linkcolor=blue, citecolor=blue]{hyperref}
\usepackage{tabularx}
\usepackage{graphicx}
\usepackage{amssymb,bm}
\usepackage{xcolor}
\usepackage{amsmath}
\usepackage[varg]{txfonts}
\usepackage{enumerate}
\usepackage[normalem]{ulem}
\usepackage{subfigure}
\usepackage[all]{hypcap}

\newcommand{\smsection}[1]{%
  \refstepcounter{section}%
  \section*{\thesection.\ #1}%
}

\newcommand{\PMO}{\affiliation{Purple Mountain Observatory, Chinese Academy of Sciences, Nanjing 210023, China}}
\newcommand{\USTC}{\affiliation{School of Astronomy and Space Sciences, University of Science and Technology of China, Hefei 230026, China}}
\newcommand{\equal}{\affiliation{Contributed equally to this work}}

\begin{document}

\title{Probing Extragalactic Ultralight Axion Dark Matter with Decade-Long Blazar Optical Polarimetry of 1ES 1959+650}

\author{Qiu-Ju Huang}\PMO\USTC\equal
\author{Bao Wang}\PMO\USTC\equal
\author{Jun-Jie Wei}\thanks{jjwei@pmo.ac.cn}\PMO\USTC
\author{Xue-Feng Wu}\thanks{xfwu@pmo.ac.cn}\PMO\USTC

\begin{abstract}
Axions or axion-like particles (ALPs) are well-motivated dark matter (DM) candidates whose 
coupling to photons induces periodic oscillations in the polarization angle of astrophysical 
light. This work reports the first search for such a signature using 10 years of optical 
polarimetric monitoring of the blazar 1ES~1959+650. No statistically significant periodicity 
is detected, determined by a Lomb-Scargle periodogram combined with Monte Carlo analysis. 
Assuming a central DM density in the host galaxy, this null result places tight upper limits 
on the ALP-photon coupling constant at
\(g_{a\gamma}<(1.3\times10^{-13}-6.0\times10^{-10})\,\mathrm{GeV}^{-1}\) across a broad ALP 
mass range of $m_a \sim (1.4\times10^{-23}-5.2\times10^{-20})\,\mathrm{eV}$. Assuming a soliton-core DM density, we obtain upper limits of \(g_{a\gamma}< (2.5\times10^{-13}-6.2\times10^{-13})\,\mathrm{GeV}^{-1}\) over the same ALP mass range. Our constraints 
surpass those from Very Long Baseline Array polarimetry of active galactic jets 
and from polarimetric data of the European Pulsar Timing Array, and are 
competitive with those from long-term Galactic pulsar timing of PSR J0437-4715 over the same 
ALP mass window. These results establish long-term blazar polarimetry as a competitive and
complementary approach for probing axion-like DM on extragalactic scales.
\end{abstract}


\maketitle

\section{Introduction}

The true nature of dark matter (DM), despite constituting a major portion of the Universe, 
remains a defining challenge in modern astrophysics and particle physics \cite{rubin1980,clowe2006,larson2011,bertone2010}. 
Because of its success in explaining the DM relic abundance, weakly interacting massive particles 
(WIMPs) have long been the leading hypothetical candidate for DM \cite{BERTONE2005}. However, 
no compelling evidence for WIMPs has been found in direct detection experiments, which have 
instead placed very stringent limits on the cross-section for DM interactions with standard 
model particles \cite{2010ARA&A..48..495F,2017NatPh..13..212L}. This persistent null result 
has spurred growing interest in searches for alternative DM candidates \cite{2018arXiv181107873S,2020ScPC....3....7B}. 

Among the many proposed DM candidates, axions and axion-like particles (ALPs) have attracted significant attention due to their strong theoretical motivation \cite{jaeckel2010low, Preskill1983PhLB, Abbott1983PhLB, Dine1983PhLB, Hui2021ARA&A, Khlopov1985, Berezhiani1991, Berezhiani1992}. The axion was originally introduced by \citet{peccei1977cp} as a solution to the strong CP problem. 
More generally, ALPs represent a broader class of light pseudoscalar bosons that emerge in many extensions of the standard model \cite{jaeckel2010low,Svrcek2006,Arvanitaki2010}. Unlike the axion, for which the mass and couplings are typically related 
by the underlying theory, ALPs can arise in effective field theories 
where their masses and photon couplings are independent \cite{jaeckel2010low,2018arXiv181107873S}.
A defining feature of ALPs is their potential to be extremely light 
($m_a \sim 10^{-22} \;{\rm eV}$), making them a canonical example of bosonic ultralight DM,
also known as fuzzy DM or wave DM \cite{Goodman2000, Hu2000, Hui2017, Niemeyer2020, Hui2021ARA&A, Peebles2000, Sakharov1994, Sakharov1996, Khlopov1999}.
At such low masses, these particles possess an extremely large de Broglie wavelength, 
resulting in a suppression of small-scale structure formation from the quantum pressure
generated by the uncertainty principle \cite{Hui2021ARA&A,Peebles2000}. This effect helps alleviate 
two key small-scale issues: the core-cusp problem and the satellite problem. It resolves 
the tension in which cold DM simulations overpredict the abundance of cuspy halo density profiles
and low-mass halos compared to observations \cite{Weinberg2015,Hu2000,Hui2017,Deng2018,Rubin2001,Khlopov2002,Khlopov2005}. 
Ultralight ALPs can be probed using various astrophysical methods, including gravitational lensing \cite{Lin2024MNRAS, Sipple2025MNRAS, Hou2026arXiv}, black-hole super-radiance \cite{Arvanitaki2011, Cardoso2018JCAP}, and polarization-based searches 
\cite{Ivanov2019,Chen2020,Liu_2020,Yuan2021,Gan_2024,Fedderke_2019,Fujita_2019,Caputo_2019,Castillo2022,Liu2023,Finelli2009PhRvD, Galaverni2023PhRvD}. Among these, black-hole super-radiance directly probes the ultralight ALP mass via interactions with spinning black holes and, in some models, does not require a sizable ALP-photon coupling \cite{Cardoso2018JCAP}. 
In contrast, probing the weak photon-ALP couplings of ultralight ALP DM with astrophysical
systems remains challenging, which makes polarization-based probes of ALP-induced birefringence particularly valuable.

ALPs couple to the electromagnetic field via the Lagrangian term 
$\mathcal{L}_{a\gamma} = \frac{1}{4} g_{a\gamma} a F_{\mu\nu}\tilde{F}^{\mu\nu}$, where $a$ is 
the ALP field, $g_{a\gamma}$ is the ALP-photon coupling constant, and $F_{\mu\nu}$ is the 
electromagnetic field tensor. This interaction alters the propagation of electromagnetic waves 
in an ALP field, producing a birefringent effect on polarization. Specifically, it induces 
opposite phase shifts in the left- and right-handed circular polarization modes due to 
modifications in their dispersion relations, a phenomenon known as cosmic birefringence 
\cite{Carroll1990,Harari1992}. For linearly polarized light, this results in a periodic 
rotation of the polarization angle (PA), with an oscillation frequency set by the ALP mass 
and an amplitude proportional to $g_{a\gamma}$. This effect enables the probing of ALPs 
through long-term polarimetric monitoring of astrophysical sources.

Several studies have exploited this mechanism to constrain the ALP parameter space
\cite{Ivanov2019,Chen2020,Liu_2020,Yuan2021,Gan_2024,Fedderke_2019,Fujita_2019,Caputo_2019,Castillo2022,Liu2023,Finelli2009PhRvD, Galaverni2023PhRvD}. 
For instance, Very Long Baseline Array (VLBA) polarimetry of parsec-scale jets 
in active galaxies has been used to search for coherent PA oscillations induced by 
ultralight axion-like DM \cite{Ivanov2019}. Likewise, the analysis of long-term 
Parkes Pulsar Timing Array data has produced constraints from Galactic pulsars 
\cite{Caputo_2019,Castillo2022,Xue2026PhRvL,Liu2023}, while 38-day polarization measurements 
of fast radio bursts (FRBs) with the Five-Hundred-Meter Aperture Spherical radio Telescope 
(FAST) have provided complementary constraints \cite{wang2025}. In all these cases, 
the nondetection of a periodic signal has placed upper limits on the ALP-photon coupling 
constant $g_{a\gamma}$ across a range of ALP masses.

Blazars are a class of active galactic nuclei (AGNs) with relativistic jets 
aligned close to the observer's line of sight \cite{Urry_1995, Padovani_2017, Saikia_2022}. 
Their radio-to-optical continuum is dominated by highly polarized synchrotron emission 
\cite{zhang2014ApJ, Fraija2017ApJ, Agudo2018MN}, making them ideal extragalactic 
laboratories for probing ALP-photon interactions. While previous ALP studies using
blazars have primarily relied on spectral analyses 
\cite{Reesman2014JCAP, Galanti2020MN, Li2022ChPhC, Galanti2023PRD, Gao2025JCAP, Zhou2025JCAP, Pant2024PhRvD}, 
their polarimetric properties have not yet been applied in this context. 
We address this gap using the decade-long optical polarization dataset of the blazar 
1ES 1959+650 \cite{Singh_2025}, which is hosted by a gas-rich elliptical galaxy 
\cite{Heidt1999A&A, Falomo1999A&A, Scarpa1999ApJ}. Theoretically, the ALP-induced 
PA oscillation period scales as $T'=2\pi(1+z)/m_a$, implying that a long observational 
baseline is essential to probe lower ALP masses across a broad range---here spanning 
nearly three orders of magnitude. Combined with the intrinsically high polarization 
fraction of blazar synchrotron emission, this dataset provides a uniquely sensitive 
and extragalactic probe for ALP signatures.

In this work, we carry out the first search for axion-like DM using 10 years of optical 
polarization observations of 1ES 1959+650. A Lomb-Scargle (LS) periodogram and Monte Carlo 
analysis of the PA time series reveal no significant periodic signal. Based on this 
nondetection, we set upper limits on the ALP-photon coupling strength at 
\(g_{a\gamma}<(1.3\times10^{-13}-6.0\times10^{-10})\,\mathrm{GeV}^{-1}\)
for ALP masses of $m_a \sim (1.4\times10^{-23}-5.2\times10^{-20})\,\mathrm{eV}$.

\section{Models and Data}\label{sec:2}
\subsection{ALP-photon Coupling}
The coupling between the ALP and electromagnetic fields can induce a periodic rotation in the PA 
of linearly polarized photons, an effect known as cosmic birefringence. This process is governed 
by the Lagrangian \cite{wilczek1987}:
\begin{equation}
    \mathcal{L}=-\frac{1}{4}F_{\mu\nu}F^{\mu\nu}+\frac{1}{2}\left(\partial_\mu a\partial^\mu a-m_a^2a^2\right)+\frac{1}{4}g_{a\gamma}aF_{\mu\nu}\tilde{F}^{\mu\nu},
    \label{eq1}
\end{equation}
where the first two terms are the standard Lagrangians for the electromagnetic field tensor 
$F_{\mu\nu}$ and a scalar ALP field $a$ of mass $m_a$, respectively. The final term describes 
the ALP-photon interaction, with $\tilde{F}^{\mu\nu}$ being the dual tensor of the electromagnetic 
tensor and $g_{a\gamma}$ the corresponding coupling constant. In natural units ($\hbar=c= 1$), 
$g_{a\gamma}$ has dimensions of inverse mass. This coupling differentially modifies the dispersion 
relations for left- and right-handed circular polarization modes, leading to a relative phase shift 
during propagation. Specifically, the frequency ($\omega_{\pm}$) of each polarization state is 
modified according to the dispersion relation \cite{Ivanov2019}
\begin{equation}
    \omega_{\pm} \simeq k \pm \frac{1}{2} g_{a\gamma} n^\mu \partial_\mu a ,
    \label{eq2}
\end{equation}
where $n^\mu$ is the null tangent vector of the photon trajectory and $k$ is the wave vector. 
For linearly polarized photons, this frequency splitting manifests as a rotation of the polarization plane. 
The net rotation accumulated along the photon's path $C$, from the source at spacetime coordinates 
$(x_1,\,t_1)$ to the observer at $(x_2,\,t_2)$, is given by
\begin{equation}
\begin{split}
    \phi(t) & = \frac{1}{2} \int_C \left(\omega_{+}-\omega_{-} \right) {\rm d s} \\ 
    & = \frac{1}{2} g_{a\gamma} \left[ a(x_2,\,t_2) - a(x_1,\,t_1) \right] .
    \label{eq3}
\end{split}
\end{equation}
The evolution of the ALP field is described by the Klein-Gordon equation.
Neglecting the friction term, the ALP field can be approximated as an oscillating scalar field as follows: 
\begin{equation}
    a(x,\,t) = a_0(x) \sin\left( m_a t + \delta \right) ,
    \label{eq4}
\end{equation}
where $m_a$ is the ALP mass, $\delta$ is a position-dependent phase, and $a_0$ denotes 
the local field amplitude. The corresponding ALP energy density is $\rho_a =  m_a^2 a_0^2 / 2$, 
which can be identified with the local DM density $\rho_{\rm DM}$ under the assumption 
that ALPs make up the dominant DM component. If the ALP density at the observer location 
is negligible compared to that near the source (i.e., $a(t_2) \ll a(t_1)$), as discussed further in
Sec.~\ref{sec:2C}, the PA rotation 
simplifies to a simple harmonic oscillatory form as follows:
\begin{equation}
    \phi(t) = - \frac{\sqrt{2}}{2} \, g_{a\gamma} \, m_a^{-1} \, \rho_{\rm DM}^{1/2} \, \alpha \, \sin\left( \frac{2\pi t}{T'} + \delta \right) ,
    \label{eq5}
\end{equation}
where $T' = T(1+z) = 2\pi(1+z) / m_a$ is the oscillation period observed on Earth, 
accounting for the cosmological redshift $z$ of the source. The stochastic factor $\alpha$, 
which models random fluctuations, follows a Rayleigh distribution, $f(\alpha)=\alpha \exp{(-\alpha^2/2 )}$ \cite{Foster2018PhRvD}.
These fluctuations are prominent when the observation timescale is much shorter than 
the ALP field's coherence time \cite{Foster2018PhRvD, Centers2021NatCo}. Equation~\eqref{eq5} 
encapsulates the key observational signature: a periodic modulation of the PA with a period 
determined by the ALP mass. Therefore, detecting or constraining such oscillations in 
long-term polarization measurements provides a probe of the ALP-photon coupling constant 
$g_{a\gamma}$ within a specific mass range.

\begin{figure*}[htbp]
	\centering
    \includegraphics[width=0.8\textwidth]{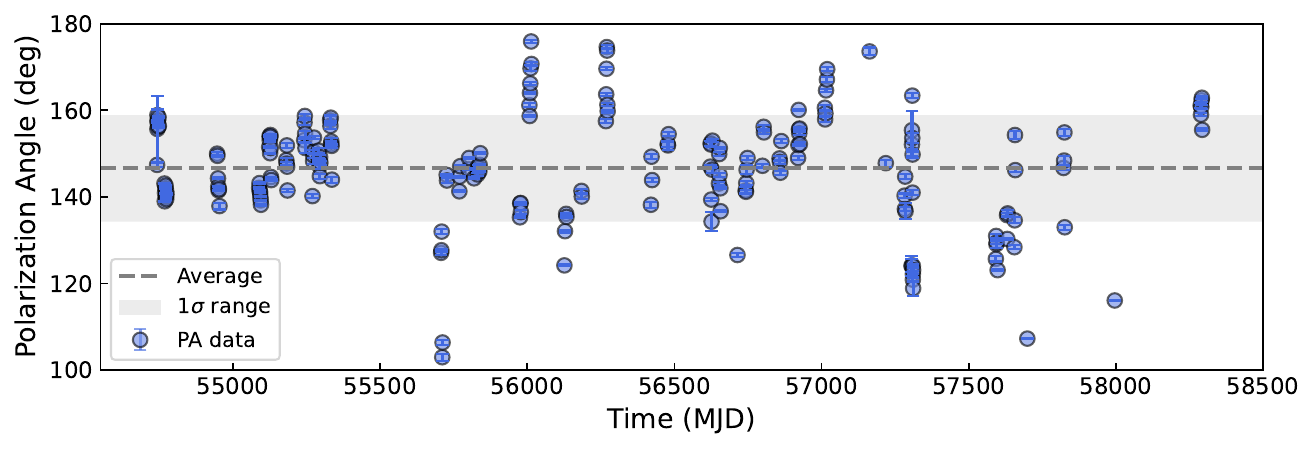}
	\caption{Long-term linear polarization measurements of the blazar 1ES 1959+650 over a decade.
	Individual PA measurements are shown as blue dots, with vertical error bars indicating 
	the $1\sigma$ uncertainties. 
	As described in \citet{Singh_2025}, the PA measurements were calibrated using interstellar polarization standard stars.
	 The reported PA uncertainties include only the statistical measurement errors, with no separate instrumental systematic uncertainty provided. The dashed line
    indicates the mean PA value of $146.61^{\circ}$, and the gray shaded band represents the
    $1\sigma$ range of the observed PA distribution, corresponding to $12.30^{\circ}$.}
    \label{fig1}
\end{figure*}

\subsection{Polarization Observations of the Blazar 1ES 1959+650}
1ES 1959+650, a well-studied extragalactic blazar at a redshift of $z\sim0.048$ 
\cite{Perlman1996ApJS, Heidt1999A&A, Falomo1999A&A}, is frequently used in studies of 
its broadband spectral energy distribution \cite{MAGIC2020A&A, Sahu2021ApJ, Ghosal2022MNRAS}. 
The stability of its optical emission and its decade-long polarization record also make it 
an ideal candidate for detecting axion-like DM. Here, we use optical linear polarization data 
of 1ES 1959+650 \cite{Singh_2025}, obtained with the Steward Observatory 
spectropolarimeter\footnote{\url{http://james.as.arizona.edu/~psmith/Fermi/DATA/individual.html}} 
in the 500- to 700-nm band, to search for ALP-induced PA oscillations and constrain the ALP-photon 
coupling constant $g_{a\gamma}$. This dataset spans nearly 10 years of observations, from October 1, 2008, 
to 30 June, 2018 (MJD 54739--58299). Fig.~\ref{fig1} shows the PA measurements as a function of time,
with a mean value of $146.61^{\circ} \pm 12.30^{\circ}$.

This polarization dataset offers two key advantages. First, it enables the first use of 
extragalactic blazar polarization to constrain the ALP-photon coupling constant, opening a new avenue 
in the search for axion-like DM. Second, the 10-year observational window allows us to 
probe an extensive ALP mass range of $m_a \sim (1.4\times 10^{-23}-5.2\times 10^{-20})\, {\rm eV}$, 
covering three orders of magnitude. This range is determined by the relation $m_a=2\pi(1+z)/T'$, 
corresponding to observed oscillation periods $T'$ from approximately one day to 10 years.

\subsection{DM Density Near the Source}\label{sec:2C}
The blazar 1ES 1959+650 is hosted by a gas-rich elliptical galaxy \cite{Heidt1999A&A, Falomo1999A&A, Scarpa1999ApJ}.
A statistical study of blazar emission regions by \citet{Fan2023} places the dominant emission region typically 
at about $\mathrm{0.076\,pc}$ from the central supermassive black hole. This location lies within the central 
AGN environment, where a high DM density is expected. Based on an analysis of lensing and kinematic data 
by \citet{Lyskova_2018}, which estimates the DM energy density in such regions to be 
$\rho_{\text{DM}} \sim  20\, \mathrm{GeV\,cm^{-3}}$,
we adopt this value as characteristic of the vicinity of 1ES 1959+650. 
We emphasize that our analysis is only mildly affected by
the uncertainty in the adopted DM density; see Secs.~\ref{SM_sec:A} and~\ref{SM_sec:B} of Supplemental Material for a detailed
discussion.
Given that this value exceeds 
the local DM density at Earth ($\simeq 0.38\,{\rm GeV\,cm}^{-3}$) \cite{Nesti2013} 
by nearly two orders of magnitude, we consider a scenario where the ALP field strength 
near the source is much greater than that in the local Milky Way, i.e., $a(t_2) \ll a(t_1)$. 

\section{Analysis Methods and Results}\label{sec:3}
The observed variations in PAs of blazars are complex, originating from multiple physical processes 
within the relativistic jet \cite{Marscher2014ApJ, Lyutikov2017MN, Zhang2019, Otero-Santos2023MN}. 
The observed PAs can be modeled as a superposition of an astrophysical background component 
($\phi_{\rm bkg}$) and a component from ALP-photon coupling [$\phi(t)$] as follows:
\begin{equation}
    \phi_{\rm obs} = \phi_{\rm bkg} + \phi(t),
    \label{eq6}
\end{equation}
where $\phi_{\rm bkg} \simeq \langle\phi_{\rm jet}\rangle + \Delta\phi_{\rm geom}(t) + \Delta\phi_{\rm turb}(t)$. 
Here, $\langle \phi_{\rm jet} \rangle$ is the long-term average PA determined by the large-scale ordered 
magnetic field in the jet, $\Delta\phi_{\rm geom}(t)$ is a slow drift from geometric and kinematic effects 
of the jet, and $\Delta\phi_{\rm turb}(t)$ represents random short-term fluctuations due to local magnetic 
disturbances and turbulence. 
Given the complexity and unpredictability of the background, we make the 
conservative assumption of neglecting its fluctuations, thereby attributing all variability to ALP-photon 
coupling to derive conservative upper limits on the coupling constant $g_{a\gamma}$ across different ALP masses.

In this work, we employ two distinct methods to probe the ALP-photon coupling. The first involves computing 
the standard deviation ($\Delta \phi$) of the ALP-induced PA shift $\phi(t)$. The second utilizes 
the LS periodogram-based Monte Carlo method.

\subsection{Standard Deviation Method}
The amplitude of the ALP-induced PA oscillations $\phi(t)$ is characterized by its standard deviation, 
$\Delta \phi \equiv \sqrt{ \left \langle \phi^2(t)  \right \rangle }$, which we adopt as the key 
observable for constraining the ALP-photon coupling. Using Eq.~(\ref{eq5}) with fiducial values of 
$m_a \sim 10^{-23}$ eV, $\rho_{\rm DM} \sim 20\, \mathrm{GeV\,cm^{-3}}$, and $\Delta \phi \sim 12^{\circ}$, 
we derive the following upper limit on the coupling constant: 
\begin{equation}
\begin{split}
    g_{a\gamma} & < 2.4 \times 10^{-13}\, \mathrm{GeV^{-1}} 
    \left( \frac{m_a}{10^{-23}\, \mathrm{eV}} \right) \\
    & \times \left( \frac{\rho_{\text{DM}}}{20\, \mathrm{GeV\,cm^{-3}}} \right)^{-1/2}
    \left( \frac{\Delta \phi}{12^{\circ}} \right).
    \label{eq7}
\end{split}
\end{equation}
Applying this constraint across the full mass range $m_a\sim(1.4\times 10^{-23}-5.2\times 10^{-20})\,{\rm eV}$ 
gives an estimated upper limit of \(g_{a\gamma}<(3.4\times10^{-13}-1.2\times10^{-9})\,\mathrm{GeV}^{-1}\).

\subsection{Lomb-Scargle Periodogram-based Monte Carlo Method}
\label{subsec:LSP}
The LS periodogram is a widely used tool for detecting periodic signals in unevenly sampled time series 
\cite{scargle1982, VanderPlas_2018}, making it particularly suited for identifying ALP-induced oscillations 
\cite{Ivanov2019, Castillo2022, wang2025}. Unlike the classical Fourier transform, the LS periodogram 
accounts for nonuniform sampling and incorporates measurement uncertainties directly into the power spectrum estimation.
A statistically significant peak in the power spectrum $P_{\text{LS}}(\nu)$ may indicate a periodic signal. 
The significance of such a peak is assessed using the false alarm probability (FAP), defined as the probability 
that a peak of equal or greater power would occur by chance in the absence of a true signal
\cite{VanderPlas_2018}.

\begin{figure}[htbp]    
    \centering    
    \includegraphics[width=0.48\textwidth]{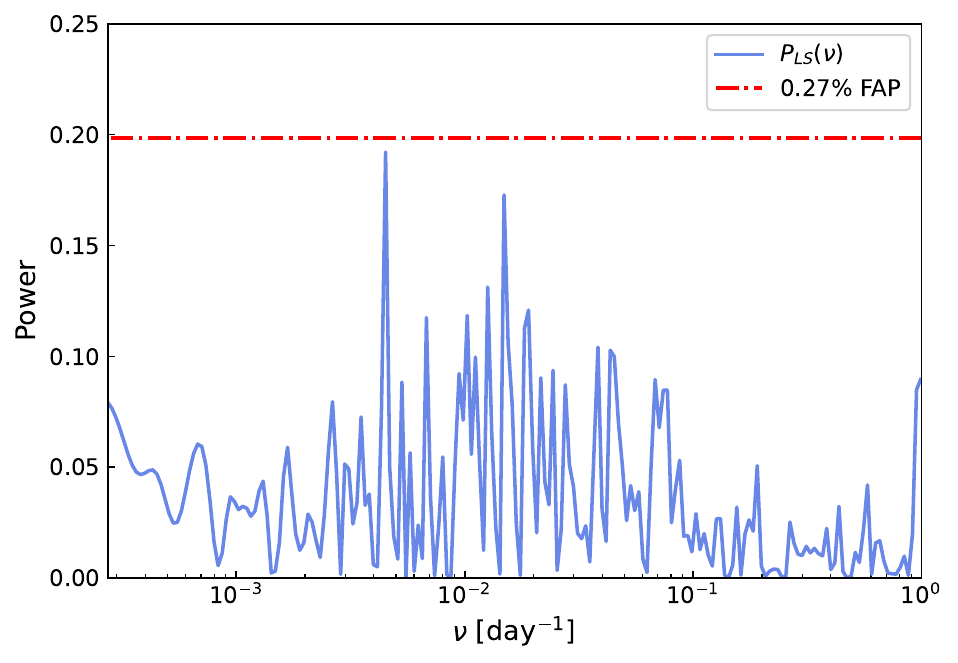}    
    \caption{LS periodogram of the PA time series for the blazar 1ES 1959+650. 
    The blue solid line shows the power spectrum ($P_{\text{LS}}(\nu)$), and 
    the red dashed line marks the FAP threshold of 0.27\%. No significant peaks 
    exceed this threshold, which is consistent with the absence of a periodic 
    signal at the 99.73\% CL ($3\sigma$).}    
    \label{fig2}
\end{figure}

We apply the LS periodogram to the optical PA data of the blazar 1ES 1959+650 using the Python package \texttt{astropy}. 
As shown in Fig.~\ref{fig2}, no significant periodic signal is detected, with all power values falling below 
the 0.27\% FAP threshold, corresponding to a $3\sigma$ confidence level (CL). In light of this null result, 
we proceed to set an upper limit on the ALP-photon coupling constant $g_{a\gamma}$.

To derive robust constraints on $g_{a\gamma}$, we employ an LS periodogram-based Monte Carlo approach 
following \citet{wang2025}. The core idea is to statistically compare the observed power spectrum 
with simulated spectra containing injected ALP-induced oscillation signals, thereby establishing 
$95\%$ CL upper limits on $g_{a\gamma}$ across a range of ALP masses. 
The simulation process consists of the following steps: 

\begin{enumerate}
    \item  We generate ensembles of synthetic background PA time series $\hat{\phi}_{\rm bkg}$ 
    by resampling the observed PA values and their measurement errors, while preserving 
    the original observation cadence.

    \item For each trial frequency $\nu_a$, we inject a periodic signal into the background time series,
    constructing the simulated data as
    \begin{equation}
      \hat{\phi}(t) = \hat{\phi}_{\rm bkg}  + \hat{\alpha} \hat{\varphi} \sin\left(2\pi \nu_a t + \hat{\delta}\right),
      \label{eq9}
    \end{equation}
    where the stochastic fluctuation $\hat{\alpha}$ is drawn from a Rayleigh distribution, 
    the amplitude $\hat{\varphi}$ is sampled uniformly from $[0, 15]$ degrees, and the phase
    $\hat{\delta}$ is uniformly distributed in $[0, 2\pi]$.

    \item For each simulated series, we compute the LS power spectrum $\hat{P}_{LS}(\nu_a)$.

    \item We identify the set of amplitudes $\hat{\varphi}$ for which the simulated power 
    $\hat{P}_{LS}(\nu_a)$ is statistically consistent with the observed power.
    From this set, we determine the value $\varphi_{95}$ such that 
    95\% of the amplitudes satisfy $\hat{\varphi}<\varphi_{95}$, representing
    the 95\% CL upper limit on the ALP-induced oscillation amplitude at frequency $\nu_a$. 

   \item  Finally, we convert $\varphi_{95}$ into an upper limit on $g_{a\gamma}$ for 
   a corresponding ALP mass $m_a$ by substituting a standard deviation of $\varphi_{95}/\sqrt{2}$ into Eq.~(\ref{eq7}).
\end{enumerate}

\begin{figure}[htbp]
    \centering
    \includegraphics[width=0.48\textwidth]{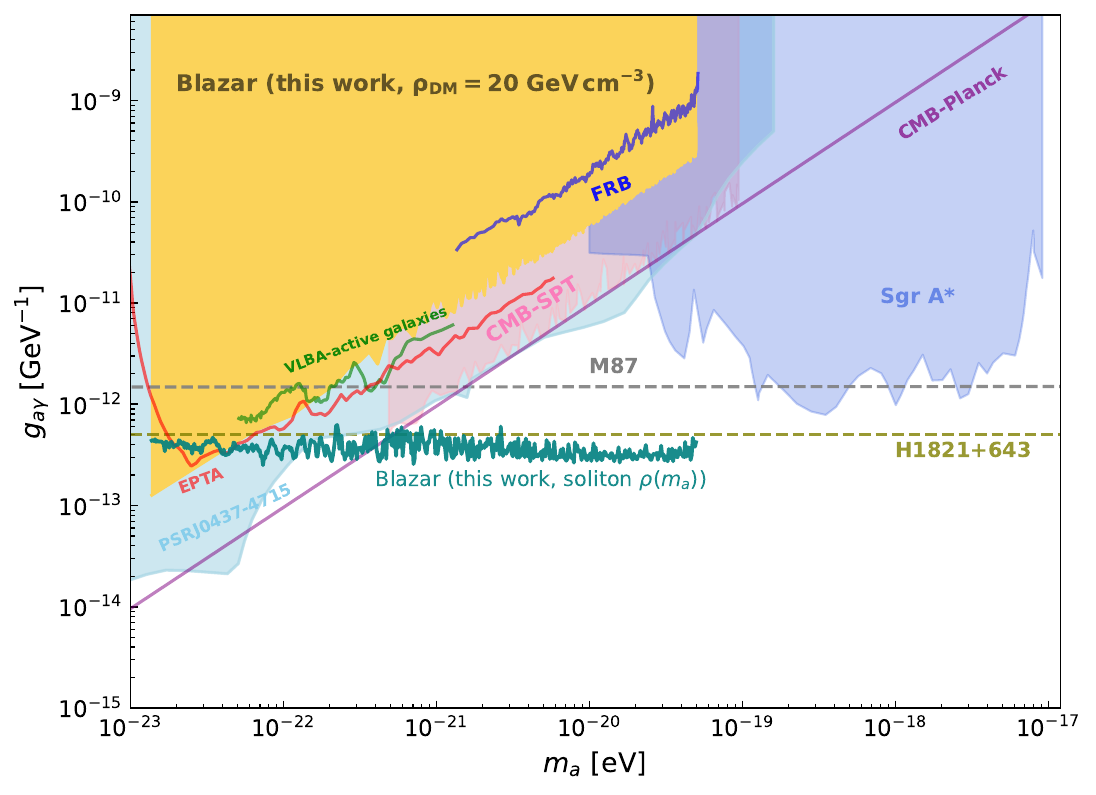}
        \caption{Constraints on the ALP-photon coupling constant $g_{a\gamma}$
    as a function of ALP mass $m_a$. The gold shaded region shows the
    95\% CL upper limit derived from 10 years of optical polarization
    monitoring of the blazar 1ES~1959+650 under the fiducial DM density assumption
    $\rho_{\rm DM}=20~{\rm GeV\,cm^{-3}}$. 
    The teal solid line shows the corresponding constraint derived from the same blazar polarimetry under the phenomenological soliton-core DM density profile.
    The comparison limits shown as solid curves or shaded
    regions are obtained under the assumption that ALPs constitute DM.
    These include 
    VLBA polarimetry of jets from active galaxies (green solid line)
    \cite{Ivanov2019}, FAST observations of FRB~20220912A (blue solid line)
    \cite{wang2025}, SPT-3G CMB data (light pink region)
    \cite{Ferguson_2022}, EHT observations of Sgr~A$^\ast$
    (royal blue region) \cite{Yuan2021}, polarimetry of
    PSR~J0437--4715 (light blue region) \cite{Caputo_2019},
    Planck CMB polarizations (purple solid line)
    \cite{Fedderke_2019}, and polarimetric data from EPTA (red solid line) \cite{Porayko2025}. 
    ALP constraints that do not require ALPs to constitute DM are shown as dashed curves, including Chandra observations of
    the quasar H1821+643 (olive dashed line) \cite{Sisk_Reyn_2021} and of
    M87 (gray dashed line) \cite{Marsh_2017}. }
    \label{fig3}
\end{figure}

This procedure is repeated for each candidate frequency $\nu_a$ in the interval $[1/3650,\, 1]\,\mathrm{day^{-1}}$, 
corresponding to the ALP mass range $m_{a}\sim[1.4\times10^{-23},\,5.2\times10^{-20}]\,\mathrm{eV}$. 
The resulting upper limits, shown as the gold shaded region in Fig.~\ref{fig3}, constrain the ALP-photon coupling constant to 
\(g_{a\gamma}<(1.3\times10^{-13}-6.0\times10^{-10})\,\mathrm{GeV}^{-1}\) across the entire mass range.

The bounds derived above depend on the assumed DM density near
the source, $\rho_{\rm DM}$, and the stochastic factor $\alpha$ in Eq.~(\ref{eq5}).
To assess their impact, we vary $\rho_{\rm DM}$ and compare two treatments of $\alpha$: a fixed value $\alpha=1$ and a stochastic value drawn from the Rayleigh distribution (see Sec.~\ref{SM_sec:A} of Supplemental Material). 
Since the ALP-induced PA oscillation amplitude is proportional to
$g_{a\gamma}\rho_{\rm DM}^{1/2}$,
the corresponding upper limit scales as $g_{a\gamma}^{\rm limit}\propto \rho_{\rm DM}^{-1/2}$.
Therefore, the uncertainty in $\rho_{\rm DM}$ affects our resulting constraints only mildly. Moreover, we demonstrate that our conclusions are not sensitive to the choice of stochastic modeling.

In addition to the fiducial DM density adopted in the main analysis, we also consider a phenomenological soliton-core DM density profile. 
For ultralight axion DM, a solitonic core with a characteristic radial density profile can form at the center of a DM halo through the balance between gravitational interaction and quantum pressure \cite{Yuan2021, Schive2014NatPh, Hui2021ARA&A, Jones2026arXiv}. We compare the constraints derived using this profile with those obtained
under the fiducial DM density assumption (see Sec.~\ref{SM_sec:B} of Supplemental Material).
Under this soliton-core DM density assumption, the derived 95\% CL upper limit is
$g_{a\gamma}<(2.5\times10^{-13}-6.2\times10^{-13})\,\mathrm{GeV}^{-1}$
over the ALP mass range
$m_a \sim (1.4\times10^{-23}-5.2\times10^{-20})\,\mathrm{eV}$ (see the teal solid line in Fig.~\ref{fig3}).
Our main conclusions remain robust at low ALP masses, whereas at higher masses the constraints become more competitive with existing polarimetric bounds.

To illustrate how injected ALP-induced PA
oscillation signals would appear for a given set of ALP parameters (mass and coupling 
strength) and to quantitatively compare them with the observed patterns in Fig.~\ref{fig2},
we construct three benchmark injected signals corresponding to long-period, intermediate-period, and short-period ALP oscillations, with periods $T_{\rm ALP}=$ 1000, 600, and 60 days, respectively (see Sec.~\ref{SM_sec:C} of Supplemental Material). For the long- and intermediate-period signals, we find that the injected ALP modulations would produce a clear enhancement of LS power relative to the observed periodogram shown in Fig.~\ref{fig2}. In contrast, the short-period signal exhibits only a moderate enhancement, due to the limited sensitivity arising from the sparse observing cadence.

\subsection{Comparison with Previous Constraints}
Fig.~\ref{fig3} compares the 95\% CL upper limits on $g_{a\gamma}$ derived in this work with those from other 
astrophysical probes. These include VLBA polarimetry of jets from active galaxies \cite{Ivanov2019}, 
FAST observations of FRB~20220912A \cite{wang2025}, SPT-3G observations of the cosmic microwave 
background (CMB) radiation \cite{Ferguson_2022}, Chandra observations of the quasar H1821+643 
\cite{Sisk_Reyn_2021} and of M87 \cite{Marsh_2017}, Event Horizon Telescope (EHT) observations of 
Sgr A* \cite{Yuan2021}, polarimetric monitoring of PSR J0437-4715 \cite{Caputo_2019},
Planck CMB polarizations \cite{Fedderke_2019}, and polarimetric data from the
European Pulsar Timing Array (EPTA) \cite{Porayko2025}.
Our constraints are more stringent than those from VLBA observations of parsec-scale jets in active galaxies \cite{Ivanov2019} and from the polarimetric data of EPTA \cite{Porayko2025},
and are competitive with the limits set by Parkes telescope observations of the Galactic pulsar PSR J0437-4715 \cite{Caputo_2019}.
These results demonstrate that optical blazar polarimetry emerges as a viable alternative for 
extragalactic axion-like DM searches, complementary to Galactic pulsar observations.

More broadly, different polarimetric methods target complementary ALP mass ranges: Pulsar Polarization Arrays probe $m_a\sim10^{-27}$--$10^{-21}\,{\rm eV}$ via pulsar correlations~\cite{Liu2023}; black-hole imaging targets $m_a\sim10^{-21}$--$10^{-16}\,{\rm eV}$ ~\cite{Gan_2024}; and our blazar polarimetry covers $m_a\sim10^{-23}$--$10^{-20}\,{\rm eV}$ in active galaxies. These distinct source classes, environments, and coherence scales make them mutually complementary.

\section{Summary and Discussion}
\label{sec:5}

In this work, we searched for signatures of axion-like DM using a decade-long optical polarimetric 
dataset of the blazar 1ES 1959+650. Our analysis is based on the phenomenon of cosmic birefringence, 
which arises from the interaction between the electromagnetic field and the oscillating ALP field.
This interaction induces a differential phase velocity between the left- and right-handed circularly 
polarized modes, leading to a periodic oscillation in the plane of linear polarization. 

We employed the LS periodogram combined with Monte Carlo simulations to search for such periodic signals 
over a wide frequency range, corresponding to ALP masses from $1.4 \times 10^{-23}\,\mathrm{eV}$ 
to $5.2 \times 10^{-20} \, \mathrm{eV}$---spanning three orders of magnitude. No statistically 
significant periodic modulation was detected. We thus derived 95\% CL upper limits on the ALP-photon 
coupling constant of \(g_{a\gamma}<(1.3\times10^{-13}-6.0\times10^{-10})\,\mathrm{GeV}^{-1}\), 
as shown in Fig.~\ref{fig3} (gold region). This decade-long polarimetric monitoring demonstrates the competitiveness of blazar observations 
for probing extragalactic axion-like DM over a broad mass range, providing a valuable complement 
to Galactic pulsar timing arrays.

In our analysis, we conservatively neglect astrophysical background fluctuations in Eq.~(\ref{eq6})
when deriving ALP-photon coupling limits. This assumption could fail to be conservative only if such fluctuations coherently cancel the ALP signal. In reality, background components (e.g., jet geometry variations, turbulent magnetic fields) are stochastic and generally uncorrelated with the ALP oscillation, mainly contributing broadband noise to the LS periodogram. A potential exception is a coherent background oscillation at exactly the same frequency; however, as shown in Sec.~\ref{SM_sec:D} of Supplemental Material, substantial cancellation would require comparable amplitudes and a nearly opposite phase---a finely tuned configuration unlikely to persist over the decade-long baseline. Hence, this does not significantly affect our conclusions.

Extending the observing baseline and improving measurement precision enhance sensitivity, enabling access to lower ALP masses and tighter coupling limits. A 20-year projection (Fig.~\ref{figS1}, Sec.~\ref{SM_sec:A} of Supplemental Material) shows that continued long-term monitoring can probe lower-frequency, longer-period ALP-induced oscillations, thereby probing lower ALP masses and improving upper limits on $g_{a\gamma}$. Although the current PA statistical uncertainties are already small, denser and more
regular temporal sampling would be particularly beneficial for detecting coherent modulations and improving sensitivity to shorter-period signals.

The sensitivity of our analysis primarily relies on the decade-long baseline, repeated optical polarimetry, intrinsic blazar polarization, and small PA uncertainties. The total baseline limits the longest probed ALP periods, while the cadence limits the shortest. LS sensitivity degrades when the trial period approaches or falls below the typical sampling interval due to poor sampling. 
To further quantify this effect, we have added an injection-recovery test in Sec.~\ref{SM_sec:E} of Supplemental Material.

Our method applies to other blazar datasets that feature long baselines, adequate cadence, significant polarization, and small PA uncertainties. In the optical band, other $\gamma$-ray blazars monitored by the Steward Observatory program---which have multiyear optical polarization data---may serve as candidate targets for similar analyses \cite{Otero-Santos2023MN}.
At other wavelengths, x-ray polarimetry with the Imaging X-Ray Polarimetry Explorer (IXPE) provides a complementary probe, as IXPE has detected x-ray polarization from 1ES~1959+650 and other blazars
\cite{Errando2024,DiGesu2023}. Current IXPE observations, with their relatively short baselines, are naturally better suited to probing shorter-period, higher-mass ALP-induced modulations. Future long-term X-ray monitoring could further
extend this approach to other wavebands and ALP mass ranges.

Our analysis assumes that ultralight ALPs constitute the full DM component responsible for the oscillating birefringence signal.
If ALPs account for only a fraction of the DM, as can occur in string theory \cite{Arvanitaki2010}, the local field amplitude would be smaller, and the limits on $g_{a\gamma}$ would weaken accordingly.

\begin{acknowledgments}
We are grateful to the anonymous referees for insightful
comments.
This work is partially supported by the National Natural Science Foundation of China 
(grant Nos. 12321003, 12422307, and 12373053), the Strategic Priority Research Program 
of the Chinese Academy of Sciences (grant No. XDB0550400), and the National Key R\&D 
Program of China (grant Nos. 2024YFA1611704 and 2024YFA1611700).
\end{acknowledgments}

\bibliographystyle{apsrev4-1}
\bibliography{references}


\section*{SUPPLEMENTAL MATERIAL}

This Supplemental Material provides additional details and validation tests that support the methods and results presented in the main Letter. 
We first examine the impact of the assumed dark matter (DM) density and of the stochastic nature of the ultralight axion-like particle (ALP) field on the derived constraints for the ALP-photon coupling. 
We then discuss an alternative phenomenological soliton-core DM density profile and compare the resulting limits with those derived under the fiducial fixed-density assumption. 
In addition, we present illustrative examples of injected ALP-induced polarization-angle (PA) modulation signals, investigate the possible effect of a coherent astrophysical background oscillation, and assess how the observing baseline and cadence affect the sensitivity of the search when employing the Lomb-Scargle (LS) periodogram.

\setcounter{section}{0}
\renewcommand{\thesection}{\Alph{section}}

\smsection{Impact of DM density and stochastic fluctuations}
\label{SM_sec:A}

\begin{figure}[htbp]
    \centering
    \includegraphics[width=0.48\textwidth]{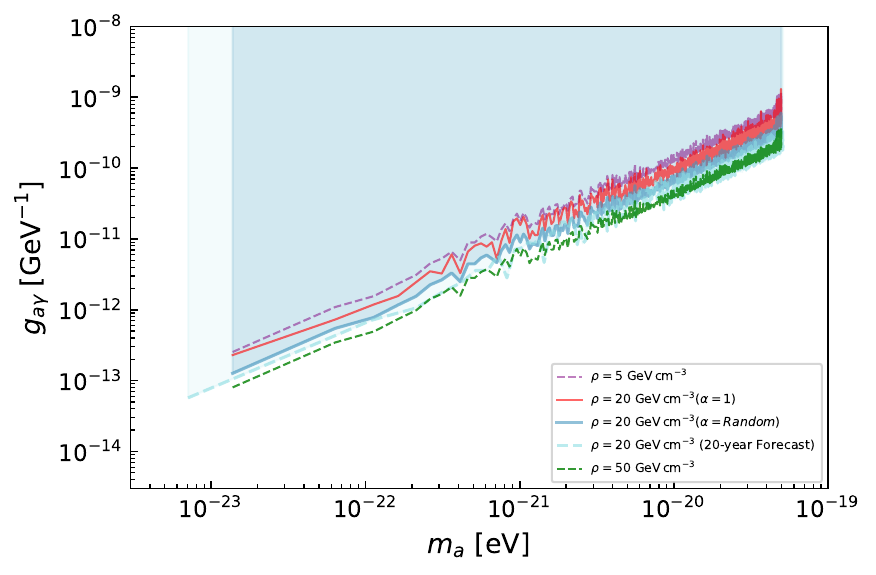}
    \caption{
     Impact of the DM density and stochastic fluctuations on the ALP-photon coupling constraints.
    The red and blue solid curves show the LS periodogram-based limits for the deterministic case ($\alpha=1$) and the stochastic case ($\alpha$ drawn from the Rayleigh distribution), respectively, both assuming the fiducial DM density $\rho_{\rm DM}=20~{\rm GeV\,cm^{-3}}$. 
    Using the same stochastic treatment for $\alpha$, the purple and green dashed curves show the limits obtained for $\rho_{\rm DM}=5$ and $50~{\rm GeV\,cm^{-3}}$, respectively. The light-blue dashed curve indicates the projected sensitivity
    for a future 20-year monitoring campaign, assuming $\rho_{\rm DM}=20~{\rm GeV\,cm^{-3}}$ and the stochastic treatment
    of $\alpha$.}

\label{figS1}
\end{figure}

The constraints derived in this work rely on several assumptions, most notably the DM energy density $\rho_{\rm DM}$ near the source. The amplitude of the ALP-induced PA oscillation scales approximately as $g_{a\gamma} m_a^{-1} \rho_{\rm DM}^{1/2} \alpha$, where $g_{a\gamma}$ is the ALP-photon coupling constant, $m_a$ is the ALP mass, and $\alpha$ accounts for the stochastic amplitude of the ultralight ALP field. Therefore, for a fixed upper limit on the oscillation amplitude, the inferred upper limit on the ALP-photon coupling approximately follows $g_{a\gamma}^{\rm limit} \propto \rho_{\rm DM}^{-1/2}$.

In the main analysis, we adopt a fiducial DM density of $\rho_{\rm DM}\simeq20~{\rm GeV\,cm^{-3}}$, based on estimates of the central DM density in elliptical galaxies hosting active nuclei \cite{Lyskova_2018}. To demonstrate the impact of this assumption, we recompute the limits for $\rho_{\rm DM}=5$ and $50~{\rm GeV\,cm^{-3}}$, as shown in Fig.~\ref{figS1}. The lower reference value, 
\(\rho_{\rm DM}=5~{\rm GeV\,cm^{-3}}\), is comparable to the central DM density inferred for NGC~1316 from planetary-nebula 
kinematics \cite{McNeilMoylan2012}. 
The higher reference value, 
\(\rho_{\rm DM}=50~{\rm GeV\,cm^{-3}}\), is considered solely to test the effect of a higher DM density assumption and is not a direct measurement; it is motivated by possible DM density enhancements induced by 
central supermassive black holes in blazar environments \cite{Lacroix2015,Ferrer2023}. Variations in $\rho_{\rm DM}$ primarily shift the derived ALP-photon coupling limits upward or downward (by a factor of approximately 2 for the lower density (\(5~{\rm GeV\,cm^{-3}}\)) and by a factor of about 1.6 for the higher density (\(50~{\rm GeV\,cm^{-3}}\))), while preserving the overall shape of their dependence on ALP mass.

We also test the effect of the stochastic amplitude of the ultralight ALP DM field. The local field is a superposition of many modes with random phases, which we model using the Rayleigh-distributed factor $\alpha$ in Eq.~(\ref{eq5}) of the main Letter \cite{Foster2018PhRvD, Centers2021NatCo}. We compare the deterministic case ($\alpha=1$) with the stochastic case, in which $\alpha$ is drawn from the Rayleigh distribution. As shown in Fig.~\ref{figS1}, both treatments yield constraints with the same overall mass dependence and exhibit only minor quantitative differences, indicating that our main conclusions are not significantly affected by the choice of stochastic modeling.

We further investigate the potential improvement offered by future long-term monitoring. Based on the current dataset, we simulate the expected sensitivity of a 20-year monitoring campaign, as shown by the light-blue dashed curve in Fig.~\ref{figS1}. Extending the observational baseline enhances sensitivity to lower-frequency, longer-period oscillations, thereby extending the accessible ALP mass range toward smaller masses. Improved PA precision or a denser observing cadence could further strengthen the limits and extend the search toward higher ALP masses. This projection highlights the long-term potential of continued optical polarimetric monitoring of blazars.

\smsection{Effect of the Alternative Soliton-core DM Density Profile}
\label{SM_sec:B}

\begin{figure}[htbp]
    \centering
    \includegraphics[width=0.48\textwidth]{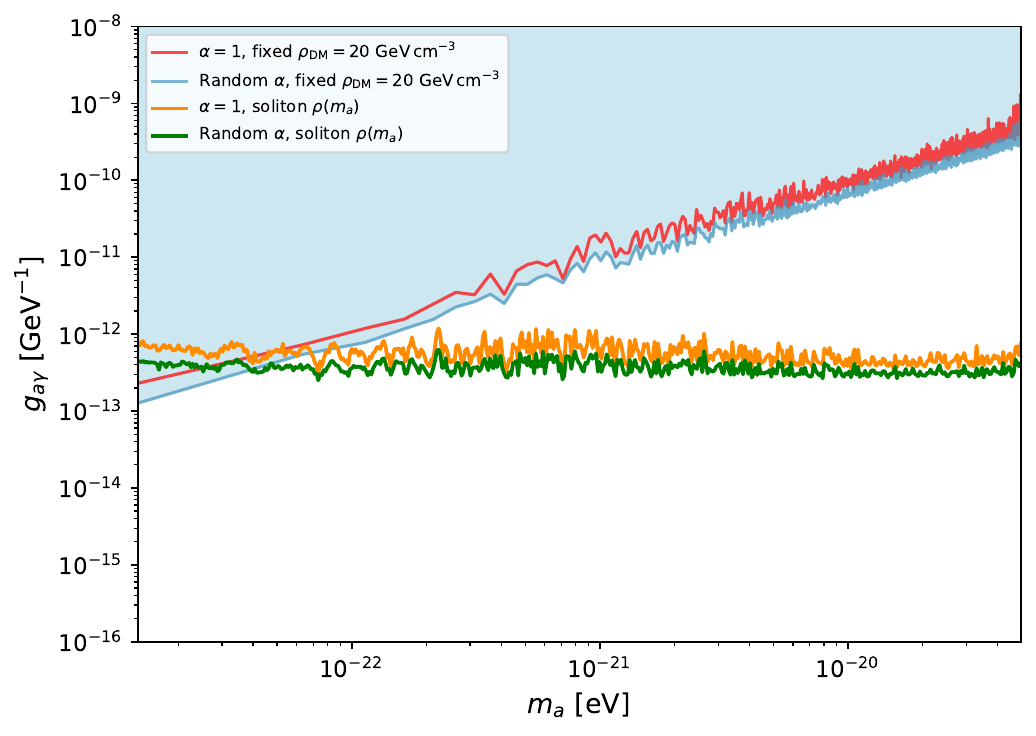}
    \caption{
     Comparison of the ALP-photon coupling constraints under different DM density assumptions. 
    The red and blue curves are obtained using the fiducial DM density adopted in the main analysis, for the deterministic case ($\alpha=1$) and the stochastic case ($\alpha$ drawn from a Rayleigh distribution), respectively. 
    The orange and green curves show the corresponding constraints derived using the phenomenological soliton-core density profile, again for the deterministic and stochastic cases, respectively.}

\label{figS2}
\end{figure}

For ultralight axion-like DM, the de Broglie wavelength can exceed the mean interparticle separation, yielding a high occupation number within a de Broglie volume. Consequently, 
the axion field can behave like a Bose-Einstein condensate, and the balance between gravity and quantum pressure can give rise to a soliton-like core near the galactic center \cite{Hui2021ARA&A}.
In addition to the fiducial DM density adopted in the main analysis, we therefore consider a soliton-core density profile to examine how a different DM environment affects the ALP-photon coupling constraints  \cite{Yuan2021,Jones2026arXiv}. 

For 1ES~1959+650, whose host galaxy is an elliptical galaxy, we adopt an estimated characteristic DM velocity scale of 
$v\simeq270~{\rm km\,s^{-1}}$, obtained by combining the host-galaxy and black-hole properties of BL Lac objects with empirical kinematic relations for early-type galaxies \cite{Falomo2003, Dutton2010MN}. 
In this scenario, the characteristic soliton-core radius is given by
\begin{equation}
    r_c = \frac{2\pi}{m_a v}.
\end{equation}
Over the ALP mass range considered in this work, $m_a\simeq(1.4\times10^{-23}-5.2\times10^{-20})~{\rm eV}$, this corresponds to 
$r_c\simeq(0.86 - 3188.16)~{\rm pc}$, which is much larger than the estimated optical polarization emission-region distance 
$r_{\rm emit}\simeq0.076~{\rm pc}$ \cite{Fan2023}. 
Hence, the emission region lies well within the soliton core over the entire mass range.

The corresponding DM density is estimated using the soliton-core profile \cite{Schive2014NatPh}:
\begin{equation}
    \rho_{\rm DM} = 0.019
    \left(\frac{m_a}{10^{-22}~{\rm eV}}\right)^{-2}
    \left(\frac{r_c}{1~{\rm kpc}}\right)^{-4}
    \,M_\odot\,{\rm pc}^{-3}.
\end{equation}
This yields a mass-dependent density spanning 
$\rho_{\rm DM}\simeq(9.4\times10^{-3} - 1.3\times10^{5})\,
M_\odot\,{\rm pc}^{-3}$, corresponding to 
$\rho_{\rm DM}\simeq (3.6\times10^{-1}$--$4.9\times10^{6})~
{\rm GeV~cm^{-3}}$ across the considered parameter space.

Using this density profile, the ALP-photon coupling constraint is rescaled relative to the fiducial fixed-density case. Because $r_c\propto m_a^{-1}$ for a fixed velocity scale, the soliton-core density scales approximately as $\rho_{\rm DM}\propto m_a^{-2}r_c^{-4}$. The coupling limit scales as $g_{a\gamma}^{\rm limit}\propto m_a\rho_{\rm DM}^{-1/2}$. Therefore, the factor $\rho_{\rm DM}^{-1/2}\propto m_a^{-1}$ approximately cancels the explicit $m_a$ dependence in $g_{a\gamma}$. As shown in Fig.~\ref{figS2}, the resulting constraint is nearly flat in the $g_{a\gamma}$--$m_a$ plane, in contrast to the increasing trend obtained under the fixed-density assumption.

We also present this soliton-core constraint in Fig.~\ref{fig3} of the main Letter, where it is compared with other existing ALP-photon coupling limits. Under this phenomenological soliton-core density scenario, the resulting 95\% CL upper limit is
$g_{a\gamma}<(2.5\times10^{-13}-6.2\times10^{-13})\,\mathrm{GeV}^{-1}$
over the ALP mass range
$m_a \sim (1.4\times10^{-23}-5.2\times10^{-20})\,\mathrm{eV}$. The constraint is nearly flat across the considered mass range and exhibits greater sensitivity than existing polarimetric bounds in parts of the parameter space. However, it should be interpreted as a complementary scenario because it relies on the phenomenological soliton-core density assumption.

\smsection{Illustrative Injection of ALP-induced PA Modulations}
\label{SM_sec:C}

\begin{figure*}[htbp]    
    \centering    
    \includegraphics[width=0.8\textwidth]{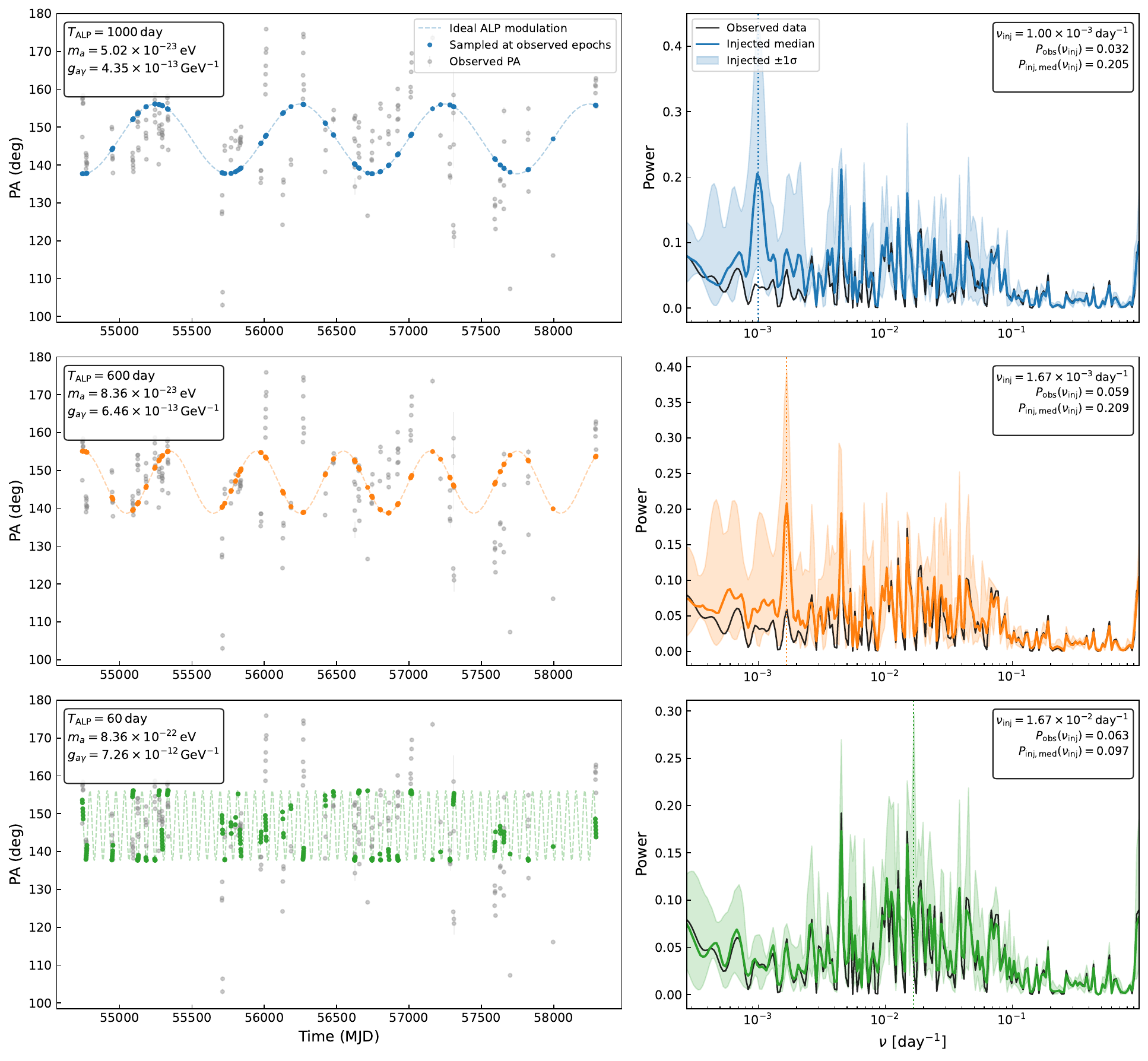} 
    \caption{
   Illustration of benchmark ALP-induced PA modulations for 1ES~1959+650.
    Left panels: PA time series for three representative ALP oscillation periods ($T_{\rm ALP}$ = 1000, 600, 60 days). Gray points show the observed PA measurements. Colored dashed lines represent the ideal sinusoidal ALP modulation computed with the corresponding ALP mass and coupling $g_{a\gamma}$ from the 95\% CL upper-limit curve. Colored points indicate the injected signal evaluated at the actual observing epochs.
   Right panels: LS periodograms of the observed PA with injected signals. The black line shows the LS power of the data. The solid colored line and shaded region represent the median and $\pm1\sigma$ range of the injected signal power, respectively. The vertical dotted line indicates the injected frequency $\nu_{\rm inj}=1/T_{\rm ALP}$.}    
    \label{figS3}
\end{figure*}

To illustrate the ALP-induced PA modulations analyzed in this work, we construct three benchmark injected signals corresponding to long-period, intermediate-period, and short-period ALP oscillations, with periods $T_{\rm ALP}=$ 1000, 600, and 60 days, respectively. For each benchmark, the ALP mass is determined by the oscillation period via
$m_a = 2\pi(1+z)/T_{\rm ALP}$, yielding $m_a = 5.02\times10^{-23}~{\rm eV}$, $8.36\times10^{-23}~{\rm eV}$, and
$8.36\times10^{-22}~{\rm eV}$ for the long-, intermediate-, and short-period cases, 
respectively. The coupling strength $g_{a\gamma}$ is taken from our derived 95\% CL 
upper-limit curve, giving $g_{a\gamma} = 4.35\times10^{-13}~{\rm GeV^{-1}}$, $6.46\times10^{-13}~{\rm GeV^{-1}}$, 
and $7.26\times10^{-12}~{\rm GeV^{-1}}$, respectively.

The injection procedure for each benchmark is as follows:

1. A random phase $\delta$ is drawn uniformly from $[0, 2\pi]$.

2. An amplitude scaling factor $\alpha$ is drawn from a Rayleigh distribution with unit scale parameter.

3. The injected PA signals at the actual observation epochs $t_i$ of 1ES~1959+650 are then computed as:
\begin{equation}
  \phi_{\rm inj}(t_i) = \phi_{\rm mean} + \alpha\,\Delta\phi_{0}\,\sin(2 \pi t_i / T_{\rm ALP} + \delta),
\end{equation}
where $\Delta\phi_{0}$ is the theoretical PA amplitude corresponding to the chosen $g_{a\gamma}$ and ALP mass, and $\phi_{\rm mean}$ is the mean of the observed PA.  

This approach preserves the overall level of the data while highlighting the injected signal's oscillatory pattern.

The left panels of Fig.~\ref{figS3} show the injected PA modulations evaluated at the actual observation epochs of 1ES~1959+650, overlaid on the observed PA measurements. The dashed colored lines represent the ideal sinusoidal ALP signals, while the colored points indicate the injected signals sampled at the observation times. These panels illustrate that lower ALP masses correspond to slower, long-period PA variations, whereas higher ALP masses correspond to faster oscillations that are more strongly affected by the sparse and irregular observing cadence. 

The right panels show the corresponding LS periodograms. The black curves represent the observed LS power, while the colored solid curves and shaded regions show the median and the $\pm 1\sigma$ range from 500 Monte Carlo realizations of the injected signals. The vertical dotted lines mark the injected frequencies. At these frequencies, the observed LS powers are $P_{\rm obs}(\nu_{\rm inj}) = 0.032$, 0.059, and 0.063, whereas the median injected LS powers are $P_{\rm inj,med}(\nu_{\rm inj}) = 0.205$, 0.209, and 0.097 for the long-, intermediate-, and short-period cases, respectively. This demonstrates that, for the long- and intermediate-period signals, the injected ALP modulations would produce a clear enhancement of LS power relative to the observed periodogram shown in Fig.~\ref{fig2} of the main Letter, while the short-period signal shows a moderate enhancement due to the limited sensitivity arising from sparse sampling.
This figure provides a visual illustration of how injected ALP-induced modulations appear in both the PA time series and the LS periodogram.

\smsection{Effect of a coherent background oscillation at the ALP frequency}
\label{SM_sec:D}

\begin{figure}[htbp]
    \centering
    \includegraphics[width=0.48\textwidth]{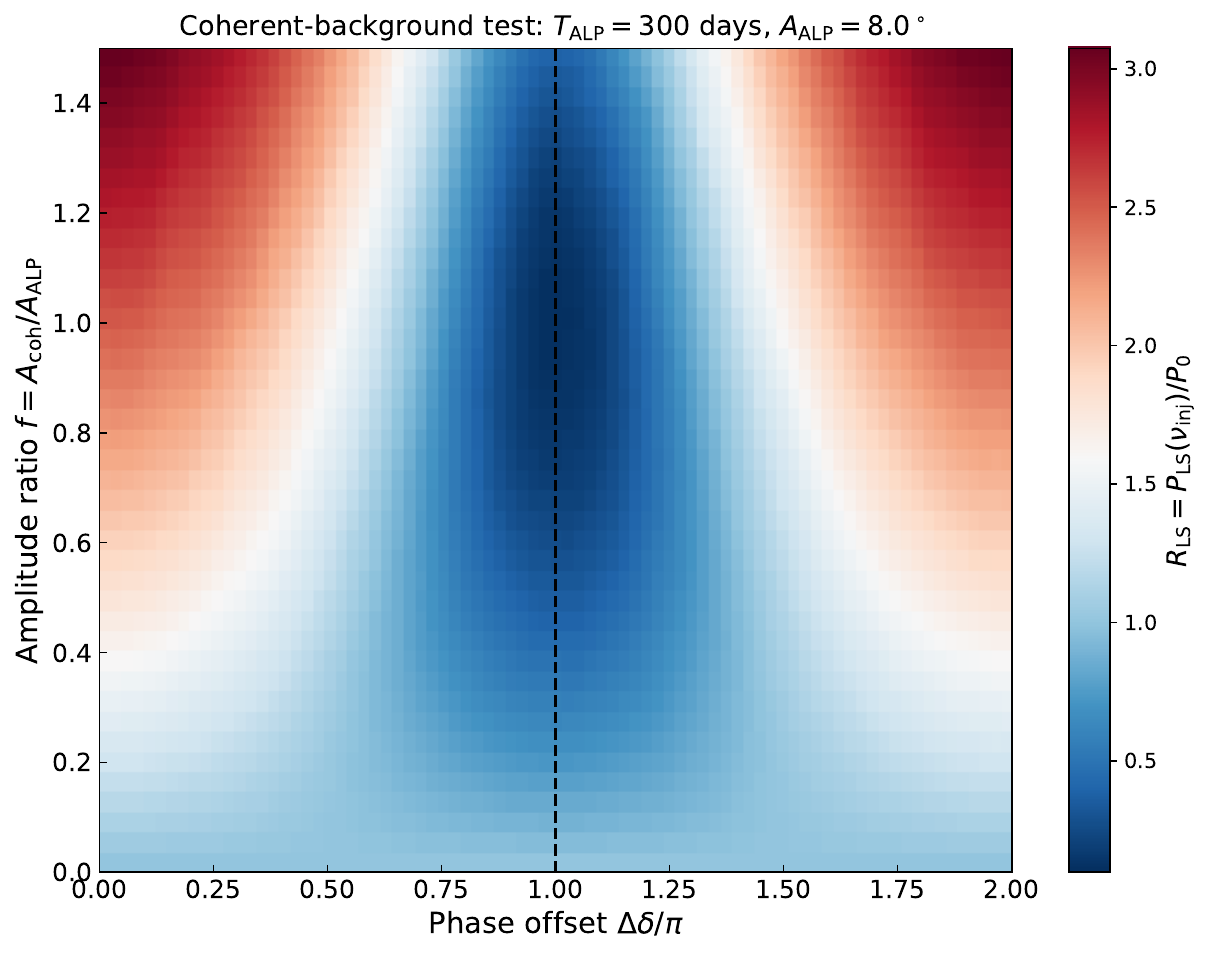}
    \caption{Effect of a coherent background oscillation at the ALP frequency. The horizontal axis shows the phase offset $\Delta\delta/\pi$ between the coherent background oscillation and the
    ALP signal, and the vertical axis shows the amplitude ratio $f=A_{\rm coh}/A_{\rm ALP}$. The color indicates the median normalized LS power at the injected frequency, $R_{\rm LS}=P_{\rm LS}(\nu_{\rm inj})/P_0$, computed over 300 Monte Carlo realizations. Here, $P_0$ is the LS power
    obtained without the coherent background component (i.e., $f=0$). The black dashed line marks the fully anti-phased case, $\Delta\delta=\pi$. The LS power at the injected ALP frequency is significantly suppressed when the coherent background has a comparable amplitude and is nearly anti-phased with the injected ALP modulation. This panel corresponds to
    $T_{\rm ALP}=300$~days and $A_{\rm ALP}=8.0^\circ$.}

\label{figS4}
\end{figure}

In the main analysis, the observed PA variations are modeled as the sum of an astrophysical background component and a possible ALP-induced periodic component. Random background fluctuations generally do not systematically affect a coherent ALP-induced oscillation; instead, they mainly contribute broadband noise to the LS periodogram. If an astrophysical periodic component
has a frequency different from that of the ALP signal, it can in principle be separated in frequency space by the LS analysis, provided the sampling is sufficient.

The limiting case is a coherent astrophysical PA oscillation at exactly the
same frequency as the putative ALP-induced signal. 
To investigate this scenario, we use simulated
PA data to quantify how such a coherent background component modifies the LS power at the injected ALP frequency. The mock PA data are generated at the actual observation epochs $t_i$ as
\begin{equation}
\begin{aligned}
\mathrm{PA}_{\mathrm{mock}}(t_i)
={}& \mathrm{PA}_{\mathrm{rand}}(t_i)
\\
&+ A_{\mathrm{ALP}}
  \sin(2\pi\nu t_i+\delta)
\\
&+ f A_{\mathrm{ALP}}
  \sin(2\pi\nu t_i+\delta+\Delta\delta).
\end{aligned}
\label{eq:mock_pa_coherent}
\end{equation}
Here, ${\rm PA}_{\rm rand}(t_i)$ is randomly drawn from the empirical
distribution of the observed PA values. The
second term describes the injected ALP-induced PA modulation, with
$A_{\rm ALP}$ being the ALP modulation amplitude, $\nu=1/T_{\rm ALP}$ the
injected frequency, and $\delta$ a random initial phase. The third term
represents a possible coherent background oscillation at the same frequency,
with amplitude expressed as $fA_{\rm ALP}$, where
$f=A_{\rm coh}/A_{\rm ALP}$ is the amplitude ratio between the coherent
background component and the ALP signal. The parameter $\Delta\delta$ denotes
the phase offset between the two sinusoidal components.

For each pair of $(f,\Delta\delta)$, we compute the LS power at the injected
frequency, $P_{\rm LS}(\nu_{\rm inj})$, and normalize it to the baseline case
without the coherent background component:
\begin{equation}
    R_{\rm LS}=\frac{P_{\rm LS}(\nu_{\rm inj})}{P_0},
    \label{eq:rls_coherent}
\end{equation}
where $P_0$ is the LS power at $\nu_{\rm inj}$ for the same simulated data but 
without the coherent background component (i.e., with $f=0$). Thus, 
$R_{\rm LS}>1$ indicates an enhancement of the LS power at the injected 
frequency, while $R_{\rm LS}<1$ indicates a suppression.

Fig.~\ref{figS4} shows the resulting map for a
representative case with $T_{\rm ALP}=300$~days and
$A_{\rm ALP}=8.0^\circ$. The adopted amplitude is chosen as a representative value comparable in magnitude to the observed PA variations, thereby providing a conservative test of signal recovery under the current sampling conditions. The strongest suppression occurs near $f\simeq1$ and $\Delta\delta\simeq\pi$, where the coherent background oscillation has a comparable amplitude and is nearly anti-phased with the injected ALP signal. 
The test demonstrates that if a coherent background oscillation partially cancels the ALP-induced modulation, the observed periodic amplitude could be reduced, potentially weakening the resulting constraint. However, substantial cancellation requires a highly specific configuration: the background oscillation would need to remain coherent over the entire observing baseline and be nearly identical in frequency and amplitude, while being opposite in phase to the ALP signal. Such a specific configuration is unlikely to persist over the decade-long observing baseline. Therefore, this potential interference is not expected to have a significant impact on the main conclusions of this work.

\smsection{Impact of Observing Baseline and Sampling Cadence on Lomb-Scargle Sensitivity}
\label{SM_sec:E}

To assess how the observing strategy affects the sensitivity of our LS periodogram search, we performed an injection-recovery test. This test is designed to illustrate how the total observing baseline and temporal sampling cadence influence the ability of the LS periodogram to recover coherent ALP-induced PA modulations.

For each mock data set, we generated simulated observing epochs with a prescribed total baseline $T_{\rm obs}$ and nominal cadence $\Delta t$. A small random time jitter was added to the nominally regular sampling to mimic mildly irregular observational cadence. The background PA values and PA uncertainties were randomly drawn from the empirical histogram distributions of the observed PA and PA uncertainty, using the same histogram-based sampling procedure as in the main injection analysis. We then injected a sinusoidal PA modulation with period $T_{\rm ALP}$, random phase, and a Rayleigh-distributed random field-amplitude factor $\alpha$, following the same treatment as in the main analysis. For this observing-strategy test, we adopt a fixed benchmark peak amplitude of $A_{\rm peak}=8^\circ$ for all injected periods. This choice allows the recovery maps to primarily reflect the effects of observing baseline and cadence.

\begin{figure*}[htbp]
    \centering
    \includegraphics[width=0.9\linewidth]{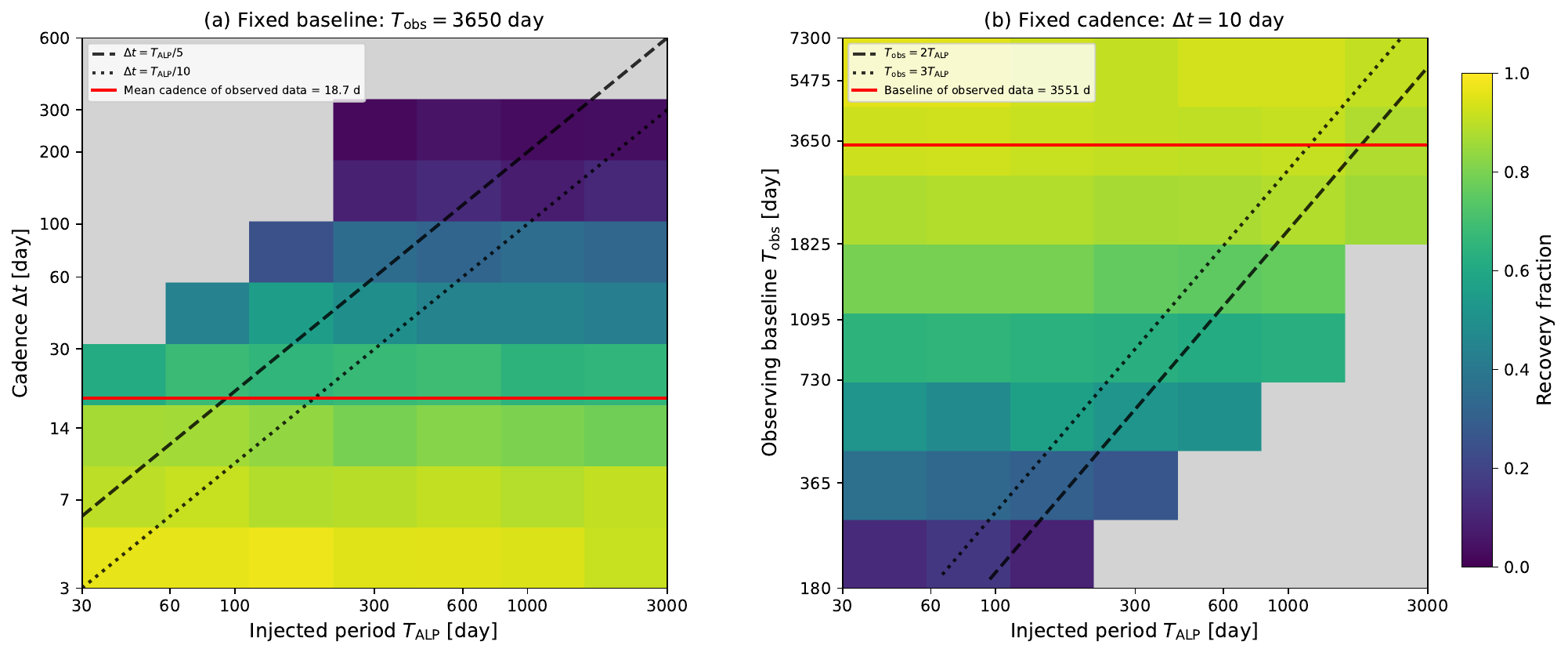}
    \caption{
    Tests of LS sensitivity against observing cadence and total observing baseline. Mock data are generated following the sampling procedure described in the main text, and a sinusoidal modulation with peak amplitude $A_{\rm peak}=8^\circ$ is injected. For each parameter set, we perform $N_{\rm MC}=300$ Monte Carlo realizations. The color indicates the recovery fraction, defined as the fraction of realizations in which the LS power within a narrow frequency range around the injected frequency exceeds the 0.27\% false-alarm threshold.
    Panel (a) shows the recovery fraction as a function of injected period $T_{\rm ALP}$ and nominal cadence $\Delta t$, with the total baseline fixed to $T_{\rm obs}=3650$ days.
    The dashed and dotted lines indicate $\Delta t=T_{\rm ALP}/5$ and $\Delta t=T_{\rm ALP}/10$, respectively, and the horizontal red line marks the mean cadence of the observed data.
    Panel (b) shows the recovery fraction as a function of injected period $T_{\rm ALP}$ and observing baseline $T_{\rm obs}$, with the nominal cadence fixed to $\Delta t=10$ days.
    The dashed and dotted lines indicate $T_{\rm obs}=2T_{\rm ALP}$ and $T_{\rm obs}=3T_{\rm ALP}$, respectively, and the horizontal red line marks the total observing baseline of the observed data.
    Gray regions denote severely undersampled configurations, for which the injected frequency lies outside the adopted LS search range or the sampling is too sparse for a meaningful recovery test.
    }
    \label{figS5}
\end{figure*}

For each mock realization, we computed the LS periodogram and examined the power near the injected frequency $\nu_{\rm inj}=1/T_{\rm ALP}$. A signal was regarded as recovered if the maximum LS power within a narrow frequency window around $\nu_{\rm inj}$ exceeded the 0.27\% false-alarm probability threshold. The recovery fraction is defined as $f_{\rm rec}=N_{\rm rec}/N_{\rm MC}$, where $N_{\rm rec}$ is the number of successfully recovered realizations and $N_{\rm MC}$ is the total number of Monte Carlo realizations.

Fig.~\ref{figS5} demonstrates that two different observational requirements govern different parts of the accessible parameter space. Long-period modulations require a sufficiently long observing baseline, while short-period modulations require sufficiently dense temporal sampling. The sampling properties of the observed data, marked by the horizontal red lines, lie in a region of relatively high recovery fraction. This indicates that, under the current observing baseline and cadence, a coherent periodic modulation with the adopted injected amplitude would be recoverable with high probability by the LS method if such a signal were present. Therefore, the absence of a significant LS peak in the observed PA time series supports interpreting the null result as a meaningful upper limit on the amplitude of such modulations.

For future applications to time-domain polarimetric data sets, a monitoring baseline spanning at least a few cycles of the target period and a cadence significantly shorter than the target period are both important. In particular, sparse sampling with $\Delta t$ approaching $T_{\rm ALP}$ would substantially reduce the sensitivity of the LS method to the corresponding ALP-induced modulation.



\clearpage

\onecolumngrid
\appendix

\end{document}